# Reply to W. Wernsdorfer's post: "*Correspondence on: Quantum interference of tunnel trajectories between states of different spin length in a dimeric molecular nanomagnet*"

**We present here an exact version of our response (dated April 27) to Wernsdorfer's correspondence submitted to Nature Physics on March 31, 2008. After consultation with a referee, Nature Physics chose not publish any part of this exchange. We would therefore like to point out that our original study has now been considered favorably by four separate referees chosen by Nature Physics. Unfortunately, Wernsdorfer subsequently posted two further variations of his correspondence on this archive [arXiv:0804.1246v1 and arXiv:0804.1246v2]. We note that aspects of the most recent posting (dated after submission of our response) contradict the version submitted to Nature Physics. However, none of the revisions add weight to Wernsdorfer's original correspondence.**

**Christopher M. Ramsey,**[1] **Enrique del Barco,**[1] **Stephen Hill,**[2] **Sonali J. Shah,**[3] **Christopher C. Beedle,**[3] **David N. Hendrickson**[3]

[1]*Department of Physics, University of Central Florida, Orlando, Florida 32765, USA.*
[2]*Department of Physics, University of Florida, Gainesville, Florida 32611, USA.*
[3]*Department of Chemistry and Biochemistry, University of California at San Diego, La Jolla, California 92093, USA.*

*To the editor* - In his Correspondence, Wernsdorfer claims several shortcomings in our recent Letter[1] concerning the observation of quantum interference involving tunneling trajectories between states of different spin length in a dimeric molecular nanomagnet. Here, we respond briefly to each of these concerns, which are shown to be either inaccurate, unjustified or, in many cases, entirely irrelevant.

It is important to first point out that Wernsdorfer's correspondence focuses on aspects of our study that are largely peripheral to the main conclusions. Indeed, he completely ignores the main evidence supporting our claims: (1) that the full pattern of hysteresis loop steps (QTM resonances) cannot be reproduced using a single-spin model, thus motivating a multi-spin description; (2) that this implies some resonances involve tunneling between states having different spin quantum numbers, e.g. $k=1$(A) and (*exc*); and (3) that the observation of magnetic oscillations in the tunneling rate found at the $k=1$(A) resonance provides evidence for quantum interference between states of different spin length. We then develop a model which reproduces these experimental findings. This model was in fact motivated by a paper published by Wernsdorfer[2] which showed that these wheel compounds should be treated as pairs of weakly coupled giant spins (dimers), with a two orders of magnitude difference between the inter- and intra-spin exchange coupling constants. This model was also confirmed in a subsequent work by Cano and collaborators[3]. In his correspondence, Wernsdorfer even confirms our assertion that a single spin description fails to explain the $k = 1$(A) resonance, and his own data can presumably (see below) also be explained in the same way.

All of the points raised in Wernsdorfer's correspondence are already known. Remarkably, our own discussion concerning the possible role of the Dzyaloshinskii-Moriya (DM) interaction, which was presented in the methods section of our letter, is copied more-or-less verbatim in Wernsdorfer's correspondence. Ironically, precisely such criticism could be leveled against Wernsdorfer's own work involving a $Mn_6$ system[4] and a $[Mn_4]_2$ dimer[5]. Although no evidence for quantum interference is presented in either of these studies, many QTM resonances are explained using multi-spin models; indeed, a simple dimer description is employed in the case of $[Mn_4]_2$. Both of these molecules possess a center of inversion preventing DM antisymmetric exchange, yet no critical discussion as to the possible interactions causing the QTM is given in these papers. Nevertheless, we would like to point out that the DM interaction used in our calculations does not substantially affect the tunnel splitting oscillations of other resonances, contrary to the claims made by Wernsdorfer. Indeed, the tunnel splittings shown in Fig. 3c (continuous lines) of our Letter were *all* calculated using a DM-interaction term. Upon reviewing the manuscript, we found a typographical error associated with the magnitude of the DM term, i.e. $\phi = 0.036$ degrees (instead of 1.5 degrees given in the paper), which corresponds to a DM interaction constant $J\sin\phi = 0.25$ mK. The effect of this interaction on the other resonances is seen only as a slight rounding of the oscillation minima.

Wernsdorfer presents his own experimental data in order to question the validity of our study, yet no information is given as to the structure of his $Mn_{12}$ sample. Our own measurements were carried out on a NEW compound that had not previously been reported in the literature. The process of synthesizing these wheels, growing high quality crystals, and performing structural and magnetic characterizations takes months. At least ten related $Mn_{12}$ wheels are known to us, and several different structures have been reported in the literature. Our own studies demonstrate how minor chemical modifications (including crystallizing from different solvents) can lead to significant variations in magnetic properties[6]. Indeed, the $Mn_{12}$ wheel first studied by Wernsdorfer[2] shows marked differences compared to the allyl wheel. Therefore, it is quite possible that the differences between Wernsdorfer's recent data and our own are due to the fact that the samples are not identical, e.g. differences in the QTM resonance fields and also the slight differences in the orientations of the magnetic axes (see below). Suffice to say, it is unclear how one can make direct comparisons without further information concerning the source of the samples, the nature of the ligand (not displayed in Fig. 1(f) and (g) in his correspondence), the solvent structure, etc.

In spite of the obvious differences, there are also similarities between Wernsdorfer's data and our own, which suggests that qualitative comparisons can be made using our model. With this in mind, we address the criticisms regarding our use of the Landau-Zener (LZ) formalism. We fully agree with Wernsdorfer's statement concerning the accuracy of the LZ method in terms of measuring tunnel splittings ($\Delta$). Indeed, we add that molecules with larger $\Delta$ relax first. Consequently, the LZ formalism really only provides information concerning a subset of the molecules, i.e. those having the largest $\Delta$. These facts are well known in the community, in part due to contributions from several authors[7] of our Letter. The main point is that the tunnel splitting oscillates as a function of the transverse field, providing evidence for quantum interference. We note that Wernsdorfer also used the LZ formalism outside of its range of applicability in his original paper on the subject,[8] yet this does not detract from the relevance of that work. Nevertheless, given the importance that Wernsdorfer places on this aspect, we wish to clarify that the measurements shown in Fig. 3a of our Letter were performed at a sweep rate of 0.4 T/min ($0.66 \times 10^{-2}$ T/s), not 0.2 T/min as assumed by Wernsdorfer. This, together with the obvious differences between our data, which could be attributed to slightly different samples, should eliminate any controversy concerning our calculations.

Finally, we wish to firmly re-state that the measurements presented in Fig. 3a of our Letter were performed by sweeping the field component applied *exactly* ($\pm 2°$) along the easy axis of the molecules, in the presence of fixed transverse fields applied *exactly* (±1°) along the hard axis of the molecules. There are 3 pairs of Jahn-Teller (JT) distorted $Mn^{III}$ ions in both the allyl version of the $Mn_{12}$ wheel and, presumably, the one studied by Wernsdorfer. The $Mn^{III}$ ions within each pair are related by an inversion. However, the pairs themselves are not related by *any* symmetry operation. Therefore, the anisotropy will be different for each pair. Consequently, it is absurd to suggest that the principal magnetic axes for the molecule can be determined in a simple way from the JT axes associated with the pairs. The only reliable way to determine the magnetic axes of such a low-symmetry SMM is through experiment, as we have done iteratively using a 3D vector magnet system (identical method to that employed by Wernsdorfer). In fact, there does not actually appear to be any difference between the hard axis directions given in Fig. 3(b) of our Letter and Fig. 1(g) of Wernsdorfer's correspondence. Furthermore, based on the way Wernsdorfer defines the plane of his molecule, it seems that we are also in reasonable agreement in terms of the easy-axis directions. Nevertheless, the smooth oscillations observed in Fig. 1c of Wernsdorfer's correspondence may well suggest a field misalignment (~5-10 degrees) in his experiment. Based on these facts and considerations, the contention that our crystal was misaligned is completely meritless.